\documentclass{jfm}
\usepackage{graphicx}
\usepackage{amsmath}
\usepackage{epstopdf,epsfig}
\usepackage[hidelinks,colorlinks=true,linkcolor=blue,citecolor=blue]{hyperref}
\usepackage{xcolor}
\usepackage{natbib}
\usepackage{newtxtext}
\usepackage{newtxmath}

\newcommand{\addc}[1]{{\color{black}#1}} 

\shortauthor{Z. Nie, J. Yao and B. Lyu}

\title{Identifying skin-friction generation structures in turbulent channel flows via  canonical correlation decomposition}

\author{Ziyi Nie, Jie Yao\aff{2}
  \corresp{Co-first author; the first two authors contribute equally.}
 \and Benshuai Lyu\aff{1}
  \corresp{\email{b.lyu@pku.edu.cn}}
 }

\affiliation{\aff{1}State Key Laboratory of Turbulence and Complex Systems, School of Mechanics and Engineering Science, Peking University, Beijing 100871, China.
\aff{2}School of Interdisciplinary Science, Beijing
Institute of Technology, Beijing, 100081, China.
}

\begin{document}

\maketitle

\begin{abstract} 
Flow structures directly responsible for local skin-friction generation in turbulent channel flows are identified using the newly developed Canonical Correlation Decomposition (CCD) method. The dominant  structures take the form of streamwise streaks that are spanwise-localised around the position where the skin-friction is targeted and exhibit significantly shorter streamwise extent than those revealed using POD. The resulting CCD spectrum shows a clear low-rank behaviour; flow reconstruction using only the first 4 CCD modes recovers more than 80\% of the examined skin friction, as opposed to 2\% recovered by the leading 4 POD modes.  When the opposition control technique is used to reduce drag,  the application of CCD shows that drag reduction is achieved by lifting the original streak structures and generating smaller streaks with opposite phases underneath. 
These findings demonstrate that CCD isolates the causally relevant flow structures governing skin-friction generation and modification,  which is expected to find use in various drag control applications in wall-bounded turbulence. 
\end{abstract}


 \vspace{-0.4cm}
\section{Introduction} 
\label{sec:intro}

Skin-friction drag in wall-bounded turbulence significantly surpasses its laminar counterpart at equivalent Reynolds numbers~\citep{kim2011physics}. In particular, it constitutes up to 50\% of the total drag on commercial aircraft, nearly 90\% for submarines, and almost all drag for internal flows (e.g., pipes/channels). Therefore, understanding the physical mechanisms responsible for skin-friction generation and developing effective drag control strategies are of central importance for energy efficiency and environmental sustainability~\citep{gad2007flow}. 

Although skin friction is inherently a wall property, defined based on the wall-normal gradient of the streamwise velocity, it
is intimately connected with flow physics across the entire wall layers. 
Over the past decades, a variety of decomposition frameworks have been proposed to relate the mean skin-friction coefficient to statistical flow quantities. For example,
the seminal contribution of \citet{fukagata2002contribution}, widely known as the FIK identity, expresses the mean skin-friction coefficient in terms of the Reynolds shear-stress distribution in wall-bounded turbulence. 
An alternative viewpoint was later provided by \citet{renard2016theoretical}, which interprets skin friction as the rate at which energy is transferred from the wall to the fluid through viscous dissipation and turbulence production.
Building on these two pioneering identities, numerous extensions have recently been developed  \citep{yoon2016contribution,de2016skin,chen2021theoretical}, significantly advancing our understanding of skin friction generation and drag-reduction mechanisms in wall-bounded flows~\citep{elnahhas2022enhancement,ricco2022integral,floryan2023fundamental,zhao2024reynolds}.

Despite the progress enabled by existing decomposition frameworks, they focus
primarily on mean or statistical quantities and therefore cannot identify the
corresponding flow structures that generate the skin friction. It is
well-established that the skin-friction drag in wall turbulence arises
predominantly from the near-wall self-sustaining cycle, namely, the
regeneration of streamwise streaks and vortices. Therefore, most
drag-reduction strategies aim to weaken or disrupt these quasi-coherent motions~\citep{rastegari2018common,gatti2016reynolds}.
At high Reynolds numbers, however, recent studies have demonstrated that
large-scale and very-large-scale outer motions play an increasingly important
role in contributing to the wall shear stress~\citep{marusic2021energy},
suggesting that skin-friction generation results from a multiscale structural
organisation spanning the entire wall layer. To probe the structural sources of
the skin friction generation, prior efforts have largely relied on
correlation-based analyses or conditional averaging techniques
~\citep{jeong1997coherent,hwang2015statistical}. Despite important insights
revealed by these methods, the connection between such flow structures and skin
friction remains largely qualitative. This underscores the need for a unified
framework that quantitatively relates coherent flow structures to skin-friction
generation, with which a more mechanistic understanding may be obtained while
consistent comparisons among different control strategies can be
made~\citep{yao2018drag}.

To achieve this goal, modal decomposition techniques may be used. Common decomposition techniques such as the Proper Orthogonal Decomposition (POD) and Direct Mode Decomposition (DMD), are widely used to extract coherent structures from complex turbulent flows.
The resulting POD modes are ranked by
their kinetic energy content, while DMD modes are ordered in terms of their
``contribution'' to the overall dynamics of the underlying system.
However, our interests are not to seek energetic or dynamically dominant structures, but rather those that contribute most directly to skin-friction generation. 
Because neither POD nor DMD is designed to target wall shear stress, a large number of modes is typically required to reconstruct the instantaneous skin friction with reasonable accuracy.
Recently, Canonical Correlation Decomposition (CCD)~\citep{Lyu2024, Lyu2024b} has been proposed as an observable-targeted decomposition technique that isolates flow structures most strongly correlated with a given quantity of interest, while suppressing uncorrelated contributions. This property makes CCD particularly well-suited for identifying the structures directly responsible for skin-friction generation.
The main objective of the present work is to apply CCD to turbulent channel flows to extract and characterise these structures. 

The paper is structured as follows. Section~\ref{sec:method}  outlines the CCD methodology and describes direct numerical simulation (DNS) databases of turbulent channel flows used. Section~\ref{sec:result} shows the resulting
flow structures that are most correlated with skin friction. The analysis is then extended to channel flows subject to opposition control, allowing the structural modifications associated with reduced drag to be examined. The following section concludes the paper.

 \vspace{-0.3cm}
\section{CCD of turbulent channel flows} \label{sec:method} 
As demonstrated by
\citet{Lyu2024, Lyu2024b}, CCD aims to decompose the flow field based on its
correlation strength with a given observable. For the sake of self-completeness,
we outline the essential steps to perform the CCD.  Consider a
sequence of  snapshots $\boldsymbol{u}_i$ obtained by sampling a flow field $u(
\boldsymbol{x}, t)$ at time $t = t_i$, where $\boldsymbol{x}$ represents the
coordinates of the flow domain, and $i$ is an integer that takes the value of
$1,2,3,\ldots, N$.
Suppose that each  snapshot is represented by a column vector of length $M$
after spatial discretisation. We write this snapshot sequence compactly in a
matrix notation as
    $\boldsymbol{U} = \left[\boldsymbol{u}_1, \boldsymbol{u}_2,
    \boldsymbol{u}_3,\ldots, \boldsymbol{u}_N\right]$.

For each $t_i$ used to sample  $\boldsymbol{u}_i$, assume that
one can simultaneously sample an observable $p(t)$  at time
$t_i+\tau_j$ and obtain the sequence $p_{i, j}$, $j=1,2,3...Q$, where $j$ is an
integer and takes the values of $1,2,3,...Q$ ($Q$ being a positive integer). In
most cases, where the observable is not significantly behind or ahead of the
flow, one can assume $\tau_1 \approx -\tau_Q/2$ for simplicity. For each integer
$i$, one defines a column vector $\boldsymbol{p}_i = [p_{i,1}, p_{i,2}, p_{i,3},
\ldots p_{i,Q}]^T,$ where $T$ denotes transpose. These vectors can be assembled into a matrix
$\boldsymbol{P}=[\boldsymbol{p}_{1}, \boldsymbol{p}_2, \ldots,
\boldsymbol{p}_{N-1}, \boldsymbol{p}_N].$

CCD starts by constructing the correlation matrix $\boldsymbol{A}$ and
performing the singular value decomposition according to
\begin{equation}
    \boldsymbol{A} = (1/\sqrt{QN^2}) \boldsymbol{P} \boldsymbol{U}^\dagger= \boldsymbol{R} \boldsymbol{\Sigma} \boldsymbol{V}^\dagger,
\end{equation}
where $\dagger$ denotes Hermitian adjoint (complex transpose), and
$\boldsymbol{\Sigma}$ is a diagonal matrix with the singular values $\sigma_j$
($j=1, 2, 3\ldots, \min(M, Q)$) as its diagonal elements. The column vectors of
$\boldsymbol{V}$ represent the CCD modes of the flow field;  these modes are mutually orthogonal, ranked by their correlation strength with the observable ($\sigma_j^2$).


To identify the flow structures most correlated with the skin friction, CCD is applied to several high-fidelity DNS databases of turbulent
channel flows, performed using the code developed by \cite{lee2015direct}. 
First, two regular uncontrolled turbulent channel flows at friction Reynolds numbers of $Re_\tau (\equiv u_\tau h/\nu) \approx  180$ and $550$, respectively, are considered. In addition, a database of opposition control with detection plane $y_d^+ = 15$ at $Re_\tau \approx 180$ \cite{yao2025drag} is included to examine how drag reduction modifies the correlation-based structures. For all cases, the computational domain size is
$4\pi h \times 2h \times 2\pi h$ in the streamwise ($x$), wall-normal ($y$) and
spanwise ($z$) directions, respectively, where $h$ denotes the
half-channel-height. 
The flow is driven by a pressure gradient, which varies in
time to ensure that the mass flux through the channel remains constant. 
\addc{The simulation time step is fixed at $\Delta t=0.01 h/U_b$ for the two databases of $Re_{\tau}\approx 180$, while it is fixed at $\Delta t=0.005h/U_b$ for $Re_{\tau}\approx 550$.  The flow snapshots are stored every 100 time steps, while the skin friction is stored every 10 time steps for all the databases.} 

For the decomposition, the streamwise velocity is chosen to represent
the flow field. The observable is the skin-friction coefficient $c_f$  at
the centre of the bottom plate, i.e. ($x$, $y$, $z$)= $(0,0,0)$. To
improve statistical convergence in the calculation of the correlation matrix
$\boldsymbol{A}$, statistical homogeneity of the flow is exploited. This is
because any point on the wall is equally suitable as an observable
location; the associated coherent structures for all observables are identical
up to a spatial shift. Thus, multiple observables may be used by spatially
shifting the velocity field relative to the observable location. For example,
for $c_f$ at $(0,0,0)$, the unshifted velocity field is used, whereas for
$c_f$ at $(\pi,0,0)$, the velocity field is circularly shifted by $\pi$ in
the streamwise direction and used as an additional sample in the ensemble space. In
this study, 192 such observables are employed, substantially improving
the statistical convergence of the resulting CCD modes.

Note that for CCD, the sampling frequency of the observable is independent of
that of the flow. Here it is sampled ten times faster \addc{for all databases} to
better resolve the temporal correlation. \addc{For example, in the uncontrolled
$Re_\tau\approx180$ case}, a total of 16870 $c_f$ samples are obtained, while
only 1687 velocity snapshots are \addc{used}. Since the correlation between the
flow and observable are likely to decay to zero beyond a time shift of $10h/U_b$, we choose $Q$
to be $100$.
Since the
temporal delay between the flow and the observable is negligible, we choose
$\tau_1$ to be $-5\Delta t_s$. For comparison, a POD analysis is also performed. 

 \vspace{-0.3cm}
\section{Results and Discussion}\label{sec:result}

In this section, the CCD analysis is applied to three DNS datasets: two
uncontrolled cases at $Re_\tau\approx180$ and $550$ respectively, and a drag-reduction
case with opposition control at $Re_\tau\approx180$. For each case, we examine the
CCD spectra and the corresponding modes that are most strongly correlated with
the instantaneous wall shear stress. The spectra and mode shapes from the POD
are also briefly presented for comparison. 



\begin{figure}
    \centering
    \includegraphics[width=0.9\textwidth]{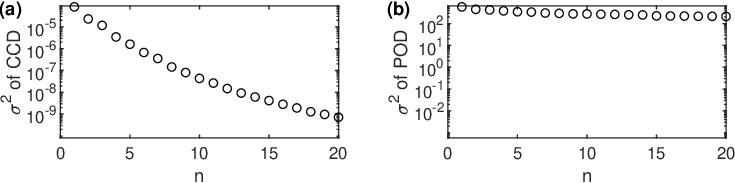}
    \caption{The spectra of (a) CCD and (b) POD modes when $Re_\tau\approx180$.}
    \label{fig:Spectra}
\end{figure}

\subsection{Uncontrolled cases at $Re_\tau\approx180$ and $550$}

Figure~\ref{fig:Spectra} shows the spectra of the first 20 CCD and POD modes for uncontrolled  $Re_\tau\approx180$ case. 
Note that for CCD modes, the singular values squared $\sigma_j^2$ measure the correlation
strength between the observable and each mode, whereas for  POD, they
reflect the kinetic energy of each mode. Compared with POD, the CCD
spectrum exhibits a much steeper decay as mode number $n$ increases, indicating
a possibility of effective order reduction. 
In particular,
figure~\ref{fig:Spectra}(a) shows that only the first few CCD modes possess significant correlation with the skin friction. By contrast, the POD spectrum remains comparatively flat, indicating that the kinetic energy of POD modes decays too slowly to yield an efficient reduced-order representation of the total flow.

\begin{figure}
    \centering
    \includegraphics[width=0.9\textwidth]{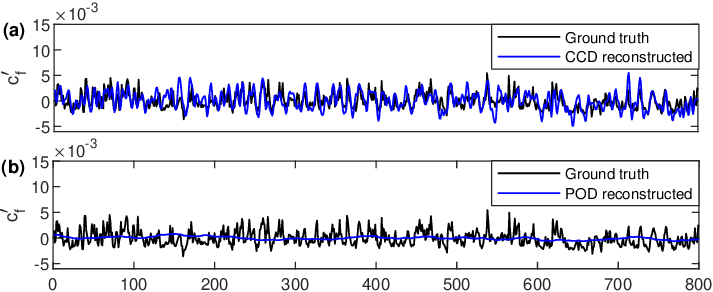}
    \caption{Reconstruction of the skin friction coefficient \addc{(fluctuation part) }using the first 4 CCD (a) and POD (b) modes when $Re_\tau\approx180$, respectively.
    }
    
    \label{fig:reconstruction}
\end{figure} 

The rapid decay of the CCD spectrum suggests that the skin friction 
may be well approximated by a low-rank model built from only a few leading CCD modes. 
Such an attempt is demonstrated in figure~\ref{fig:reconstruction}, which shows reconstructions of the instantaneous skin-friction coefficient $c_f$ obtained by projecting the flow field onto the first four CCD or POD modes. \addc{Specifically, for flow field $\boldsymbol{u}_i$ at time $t_i$, its projection on $k$-th order mode $\boldsymbol{v}_k$ is calculated by $a_k(t_i)=\boldsymbol{u}_i\boldsymbol{v}_k$. Thus, an approximation of $\boldsymbol{u}_i$ is obtained using the linear combination of the first 4 modes, i.e. $\boldsymbol{u}'_i=\sum^{4}_{k=1}a_k(t_i)\boldsymbol{v}_k$.
}

The skin friction coefficient
$c_f$ is then reconstructed using Newton's law by calculating the wall-normal gradient of
the reconstructed streamwise velocity at the observer point. 
Figure~\ref{fig:reconstruction}(a) shows that the reconstruction based on the first four CCD modes closely follows the ground truth. 
More specifically, the reconstruction
recovers more than 80$\%$ of $c_f$ variation under the $L_2$ norm. 
As the sampling frequency of the observable increases, an even higher accuracy may be expected. In contrast, the reconstruction using the first four POD modes
(figure~\ref{fig:reconstruction}(b)) performs poorly. 
This is anticipated, as since POD ranks structures by kinetic energy rather than by their contribution to $c_f$ variations. 
In summary,  figures~\ref{fig:Spectra} and~\ref{fig:reconstruction} demonstrate that CCD provides a markedly more compact representation of the $c_f$-producing structures compared with PODs.

\begin{figure}
    \centering
    \includegraphics[width=0.95\textwidth]{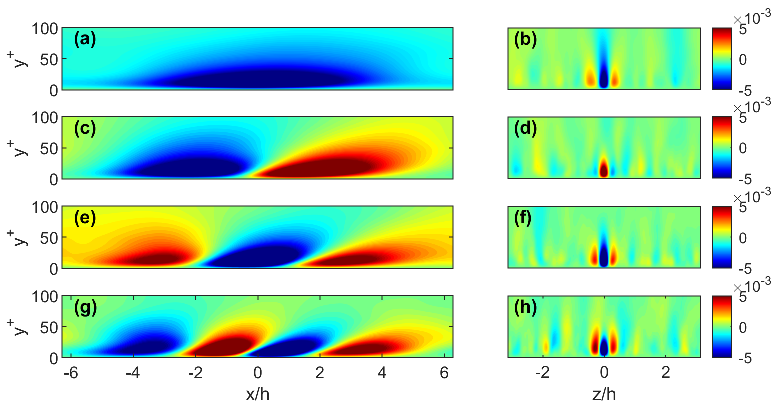} 
    \caption{Front (a,c,e,g) and left (b, d, f, h) views of the first four CCD modes when $Re_\tau\approx180$. $z=0$ for (a, c, e, g) while $x=0$ for (b, f) and $x=1/3\pi$ for (c, h).}
    \label{fig:pcdMode}
\end{figure}

\begin{figure}
    \centering
    \includegraphics[width=0.95\textwidth]{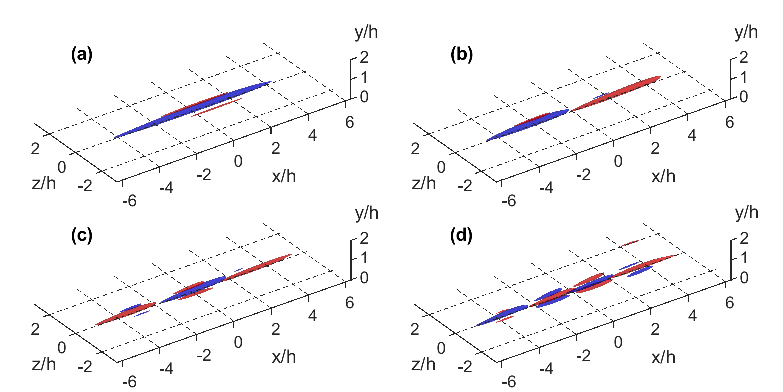} 
    \caption{Isosurface of the 1st- to 4th- order (a to d) CCD modes when $Re_\tau\approx180$. The red and blue isosurfaces represent $50\%$ and $-50\%$ of the fluctuation amplitude, respectively.} 
    \label{fig:ccdMode_3d}
\end{figure}

\begin{figure}
    \centering
    \includegraphics[width=0.95\textwidth]{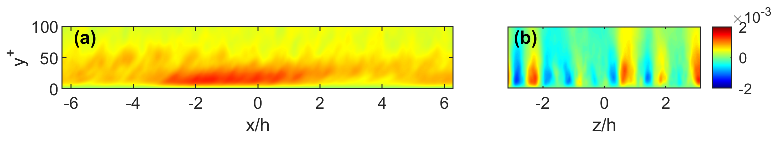} 
    \caption{Front (a) and left (b) views of the first POD mode when $Re_\tau\approx180$. $z=0.6$ for (a) while $x=0$ for (b).
    }
    \label{fig:PodMode}
\end{figure}

We are now in a position to examine the spatial structures of the resulting CCD and POD
modes. They are shown in figures~\ref{fig:pcdMode} to~\ref{fig:PodMode},
respectively. The two decompositions yield significantly different structures in both location and organisation.

Figures~\ref{fig:pcdMode}(a,c,e,g) show the front ($x$-$y$ plane) views of the
first four CCD modes. The CCD modes take the form of streamwise-elongated
streaks whose centres are aligned with the observable location on the bottom
wall. These streaks appear at a finite distance above the wall, whose centres are located at approximately $y^+\approx 15$. This height is consistent
with the well-established position of the buffer-layer streaks that dominate
momentum transfer and near-wall shear production in canonical wall turbulence.
The leading-order CCD mode exhibits the largest streamwise coherence, whereas
higher-order modes display progressively shorter characteristic wavelengths, as
manifested by the increasingly rapid phase variations. The streaks are also
slightly inclined in the $x$-$y$ plane, a feature commonly associated with the
growth and lifting of coherent packets or hairpin-like vortical structures.

Figures~\ref{fig:pcdMode}(b,d,f,h) provide the corresponding side ($y$-$z$ plane) view of the first four CCD modes. 
In the spanwise direction, the CCD modes are strongly localised around the observable point at $z=0$. 
Their effective spanwise width is approximately $0.2h$, which offers a representative estimate for the  spanwise  scale of the streaks.
This localisation indicates that only streaks sufficiently close to the observable point contribute meaningfully to the instantaneous skin friction, whereas the majority of remote structures produce a negligible contribution.
Note that although the decomposition is performed using an observable at a single wall location, this does not reduce generality because the underlying flow is statistically homogeneous in both $x$ and $z$. 

To better show the structures of the resulting CCD modes, three-dimensional isosurface visualisations are shown in figure~\ref{fig:ccdMode_3d}. These renderings reinforce the picture obtained from the planar views: the dominant CCD structures manifest as localised, streamwise-elongated streaks anchored to the vicinity of the observable point on the wall.
In addition, for streaks aligned with the channel centreline, a symmetric pair of structures is observed on either side, each with comparable scale but opposite phase. This pairing behaviour is consistent with the reflectional symmetry of the channel geometry and with the tendency of near-wall streaks to form in counter-rotating or phase-opposed arrangements as part of the self-sustaining cycle. Such paired streaks have been associated with the flanking motions induced by quasi-streamwise vortices, which naturally generate alternating high- and low-speed regions.


In contrast, the first POD mode shown in figure~\ref{fig:PodMode} also appears as a streamwise-elongated streak, but it is neither centred around the observer location nor localised in the spanwise direction. This behaviour is expected because POD prioritises structures that maximise kinetic-energy, rather than those most strongly associated with $c_f$ variations. The 2nd- to the 4th-order POD modes are similar to the first modes hence omitted.


\begin{figure}
    \centering
    \includegraphics[width=\textwidth]{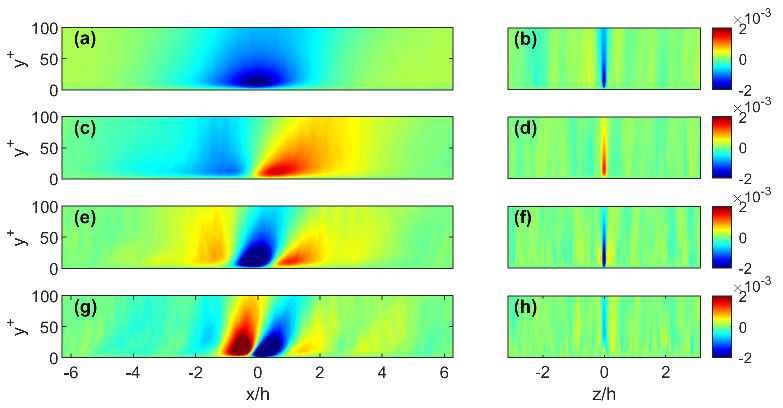} 
    \caption{Front (a,c,e,g) and left (b, d, f, h) views of the first four CCD modes when $Re_\tau\approx550$. $z=0$ for (a, c, e, g) while $x=0$ for (b, f) and $x=1/3\pi$ for (c, h).
    }
    \label{fig:ccdMode550}
\end{figure}

\begin{figure}
    \centering
    \includegraphics[width=0.95\textwidth]{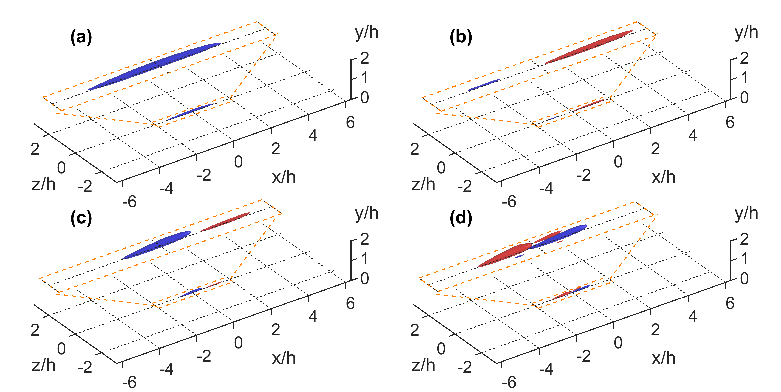} 
    \caption{Isosurface of the 1st- to 4th-order (a to d) CCD modes when $Re_\tau\approx550$. The red and blue isosurfaces represent $50\%$ and $-50\%$ of the fluctuation amplitude, respectively. Structures are zoomed-in in orange boxes ($-2<x<2$, $-0.25<z<0.25$).
    }
    \label{fig:ccdMode550_3d}
\end{figure}

As mentioned in section~\ref{sec:intro}, the skin-friction-producing structures
are also known to vary with Reynolds number.
To assess these changes, CCD is applied to a turbulent channel flow at a higher friction Reynolds number, $Re_\tau\approx550$. 
The resulting first four leading-order CCD modes are shown in
 figure~\ref{fig:ccdMode550}.
Compared with $Re_\tau\approx180$ case, the mode shapes retain a broadly similar organization and spatial distribution, suggesting that the underlying mechanism of skin-friction production remains qualitatively unchanged. Nevertheless, notable differences also exist. In particular, the streamwise extent of the structures becomes significantly shorter, consistent with the well-established trend of decreasing characteristic length scales as $Re_\tau$ increases.

\begin{figure}
    \centering
    \includegraphics[width=\textwidth]{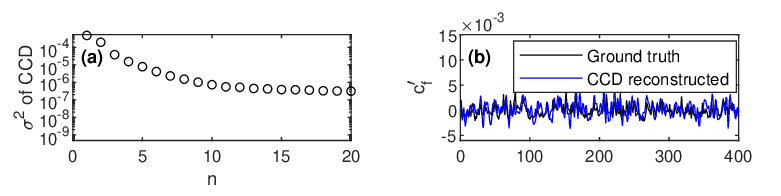}
    \caption{(a) The CCD spectra and (b) reconstruction of $c_f$ for the $Re_\tau\approx550$ case.
    } 
    \label{fig:mix}
\end{figure}

The 3D visualisations
in figure \ref{fig:ccdMode550_3d} further demonstrate that the CCD
structures decrease in size both in $x$ and $z$ directions. A key difference, evident in figure~\ref{fig:ccdMode550_3d}(a), is that the pair of opposite-phase streaks flanking the primary streak becomes much weaker compared with figure~\ref{fig:ccdMode_3d}(a), suggesting that at higher Reynolds numbers, the streaks
are spanwisely less uncorrelated.
To evaluate whether low-rank behaviour persists at higher Reynolds numbers, the skin friction is reconstructed using the leading CCD modes.
The CCD spectra and reconstruction results are shown in figure~\ref{fig:mix}.
Although the decay of $\sigma_j^2$ is somewhat slower than in the $Re_\tau\approx180$ case, the first few singular values still decrease rapidly, indicating a possibility of low-rank behaviour. Consequently, reconstruction of the instantaneous skin friction using only the first four CCD modes remains effective, as demonstrated in figure~\ref{fig:mix}(b). These results confirm that the leading CCD modes at $Re_\tau\approx550$ continue to represent the dominant flow features responsible for skin-friction production.

\begin{figure}
    \centering
    \includegraphics[width=0.9\textwidth]{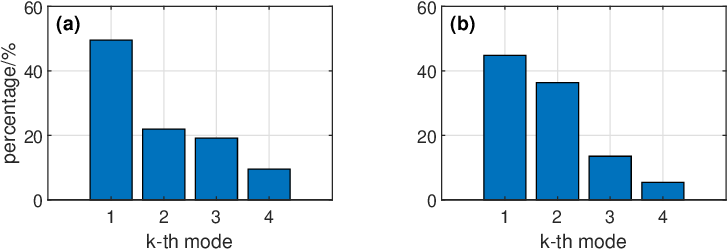}
    \caption{Modal contribution in flow reconstruction, when (a) $Re_\tau\approx180$ and (b) $550$.}
    \label{fig:histogram}
\end{figure}

\addc{To measure the contribution of each CCD mode in the skin friction
reconstruction when $Re_\tau$ varies, we define the contribution ratio of the $k$-th
order mode as $\overline{a^2_k(t)}/\sum^{4}_{n=1}\overline{a^2_n(t)}$,
and the results are shown in figure~\ref{fig:histogram}. For both
$Re_\tau\approx180$ and 550, the first-order mode accounts for more than 40\% of
the reconstructed skin friction, indicating its leading role. As $Re_\tau$
increases, the second-order mode appears to exhibit a more pronounced effect. It
is worth noting that the sampling frequency of the observable should be
increased for a better mode convergence and hence a more conclusive
demonstration of this trend at $Re_{\tau}\approx550$.}

\begin{figure}
    \centering
    \includegraphics[width=0.96\textwidth]{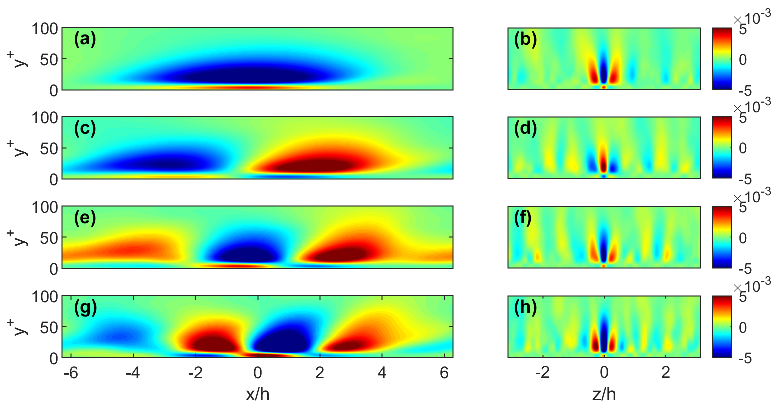} 
    \caption{Front (a,c,e,g) and left (b, d, f, h) views of the first four CCD modes for the controlled channel flow. $z=0$ for (a, c, e, g), while $x=0$ for (b, f) and $x=1/3\pi$ for (c, h).}
    \label{fig:ccdmodes_controlled}
\end{figure}

\begin{figure}
    \centering
    \includegraphics[width=0.95\textwidth]{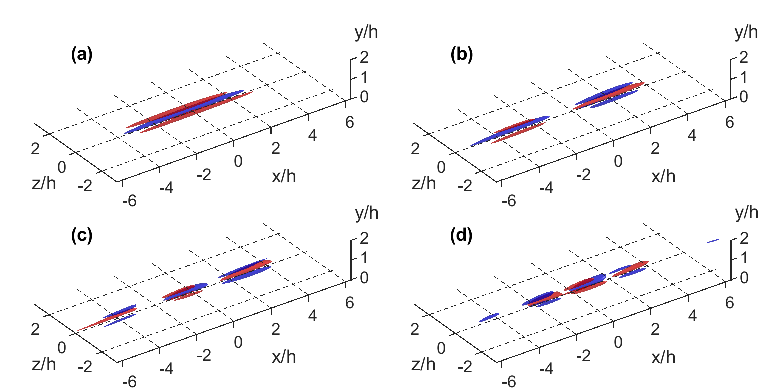} 
    \caption{Isosurface of the 1st- to 4th-order (a-d) CCD modes for controlled flow. The red and blue isosurfaces represent $50\%$ and $-50\%$ of the fluctuation amplitude, respectively.}
    \label{fig:ccdModecontrol_3d}
\end{figure}

\subsection{Case of $Re_\tau\approx180$ with opposition controls}
To examine the mechanism of opposition control, we
apply the CCD to the controlled database at $Re_\tau\approx180$.
An examination of the mean skin friction shows that a total reduction of 22$\%$ is achieved. 
The first four CCD modes for this case are shown in figures~\ref{fig:ccdmodes_controlled}–\ref{fig:ccdModecontrol_3d}. Similar to the uncontrolled cases, the dominant structures retain a streaky characteristics; however, their spatial organisation undergoes substantial modification.
Most notably, the streaks are lifted farther away from the wall and show a pronounced reduction in streamwise extent. For the leading-order mode, the streamwise length of the primary streak reduces from approximately $12h$ in the uncontrolled case (figure~\ref{fig:pcdMode}(a)) to about $8h$ under control (figure~\ref{fig:ccdmodes_controlled}(a)).
A more significant structural change is the emergence of a secondary streak immediately beneath the primary streak, as seen in figure~\ref{fig:ccdmodes_controlled}. This secondary structure is thinner, shorter, and more compact—approximately half the streamwise length and one-third the spanwise width of the primary streak—and exhibits an opposite phase.

The 3D visualisations in figure~\ref{fig:ccdModecontrol_3d}
further highlight the reorganisation of flow structures induced by opposition control.  The streak system appears more interwoven and segmented than in the uncontrolled case. In particular, the two lateral streaks flanking the primary streak become substantially more prominent—comparable in amplitude and spanwise width to the primary structure—whereas in the uncontrolled flow, they are noticeably weaker. This enhanced presence of side streaks, together with the lifted and shortened primary structure, indicates a redistribution of near-wall coherence that is consistent with the observed reduction in wall friction.



 \vspace{-0.3cm}
\section{Conclusion} 

The new flow field decomposition method, Canonical Correlation Decomposition (CCD), is applied to turbulent channel flows to identify coherent structures most strongly associated with skin friction ($c_f$) generation. 
At $Re_\tau\approx180$, $c_f$ is found to correlate primarily with a set of streamwise-elongated streaks localised in the immediate spanwise vicinity of the observable point. The CCD spectrum exhibits a pronounced low-rank character: the first four 
modes capture more than 80\% of the $c_f$ variation, in stark contrast to the slow reconstruction using the POD modes. 

When $Re_\tau$ increases to $550$, the CCD modes retain the same qualitative streaky structures but contract in both streamwise and spanwise directions, consistent with the decrease of characteristic near-wall length scales at higher Reynolds numbers. Despite these quantitative differences, the underlying mechanism associated with streak-driven wall-shear production appears largely unchanged.
The opposition-controlled case reveals a markedly different structural organisation. The primary streaks are lifted up from the wall and become shorter in the streamwise direction. More importantly, a secondary, thinner, and more compact streak of opposite phase emerges beneath each lifted streak. 
The three-dimensional visualisations further show a greater interweaving of streaks and enhanced prominence of flanking streaks, indicating a substantial reorganisation of near-wall coherent structures under control. 

In summary, these results demonstrate that CCD provides a compact and physically interpretable framework for systematically isolating the flow structures responsible for the skin-friction generation and for quantifying how these structures change under various control techniques. The structural insights obtained here offer a promising route toward an improved understanding of drag-producing mechanisms and may inform the development of more effective drag-reduction techniques in wall-bounded turbulence in future works.




\backsection[Funding]{The authors wish to gratefully acknowledge the National
Natural Science Foundation of China (NSFC) under grant numbers 12472263,
U25700222.}
\backsection[Declaration of interests]{The authors report no conflict of interest.}

\vspace{-0.4cm}
 
\bibliography{cleanref}
\bibliographystyle{jfm}

\end{document}